\documentclass{ws}

\begin{document}

\title{Calculations of single-inclusive cross sections and spin asymmetries 
in pp scattering\footnote{Talk
presented at the ``16$^{\mathrm{th}}$ International Spin Physics 
Symposium (spin2004)'', Trieste, Italy, October 10-16, 2004.}}

\author{WERNER VOGELSANG}

\address{Physics Department and RIKEN-BNL Research Center, \\
Brookhaven National Laboratory, Upton, NY 11973, U.S.A.\\
E-mail: wvogelsang@bnl.gov \\[5mm]} 

\maketitle

\vspace*{-10cm}
\begin{flushright}
BNL-NT-04/40 \\
RBRC-476 
\end{flushright}
\vspace*{8.cm}

\abstracts{
We present calculations of cross sections and spin 
asymmetries in single-inclusive reactions in $pp$ scattering.
We discuss next-to-leading order predictions as well as  
all-order soft-gluon threshold resummations.}

\section{Introduction}
Single-inclusive reactions in $pp$ scattering, such as 
$pp\to \gamma X$, $pp\to\pi X$, $pp\to {\mathrm{jet}}\,X$,
play an important role in QCD.
At sufficiently large produced transverse momentum, $p_T$, 
QCD perturbation theory (pQCD) can be used to derive
predictions for these reactions. Since high $p_T$ implies
large momentum transfer, the cross section may be factorized
at leading power in $p_T$ into convolutions of long-distance 
pieces representing the structure of the initial hadrons, 
and parts that are short-distance and describe the hard 
interactions of the partons. The long-distance contributions 
are universal, that is, they are the same in any inelastic reaction, 
whereas the short-distance pieces depend only large scales and, 
therefore, can be evaluated using QCD perturbation theory. 
Because of this, single-inclusive cross sections offer 
unique possibilities to probe the structure of the initial hadrons
in ways that are complementary to deeply-inelastic
scattering. At the same time, they test the perturbative framework, 
for example, the relevance of higher orders in the perturbative 
expansion and of power-suppressed contributions to the cross 
section. 

Of special interest is the case when the initial protons
are polarized. At RHIC, one measures spin asymmetries for
single-inclusive reactions, in order to investigate the spin structure 
of the nucleon~\cite{ref:rhic}. A particular focus here is on 
the gluon polarization in the nucleon, $\Delta g \equiv g^{\uparrow}-
g^{\downarrow}$.

In the following, we will present some theoretical predictions
for cross sections and spin asymmetries for single-inclusive 
reactions. We will first discuss the double-longitudinal spin 
asymmetries $A_{\mathrm{LL}}$ for pion and jet production at 
RHIC and their sensitivities to $\Delta g$~\cite{jssv,jsv}. 
In the second part, we will give results for new calculations~\cite{ddfv} 
of the unpolarized cross section for $pp\to\pi^0 X$ in the fixed-target 
regime, which show a greatly improved description of the available 
experimental data.

\section{Spin asymmetries for $pp\to(\pi^0,\,{\mathrm{jet}})\, X$ 
at RHIC}

We consider the double-spin asymmetry 
\begin{equation}
\label{eq:all}
A_{\mathrm{LL}} \equiv 
\frac{\sigma^{++}-\sigma^{+-}}{\sigma^{++}+\sigma^{+-}}\equiv
\frac{d\Delta\sigma}{d\sigma}\; ,
\end{equation}
where the superscripts denote the helicities of the initial 
protons. According to the factorization theorem the spin-dependent 
cross section $\Delta \sigma$ can be written  in terms of the spin-dependent
parton distributions $\Delta f$ as
\begin{eqnarray}
\label{eq:xsecdef}
\frac{d\Delta\sigma}{dp_Td\eta} 
&=& \sum_{a,b} \Delta f_a(x_a,\mu) \otimes \Delta f_b(x_b,\mu) 
\otimes \frac{d\Delta \hat{\sigma}_{ab}}{dp_Td\eta} (x_a,x_b,p_T,\eta,\mu)
\label{eq:xsecfact} \;, 
\end{eqnarray}
where the symbols $\otimes$ denote convolutions and where 
the sum is over all contributing partonic channels. We have
written Eq.~(\ref{eq:xsecfact}) for the case of jet production;
for pion production there is an additional convolution with a 
pion fragmentation function. As mentioned above, the parton-level 
cross sections may be evaluated in QCD perturbation theory:
\begin{equation}
\label{eq:partxsec}
d\Delta\hat{\sigma}_{ab}=
d\Delta\hat{\sigma}_{ab}^{(0)} + \frac{\alpha_s}{\pi}
d\Delta\hat{\sigma}_{ab}^{(1)} + \;\ldots\;\;\;,
\end{equation}
corresponding to ``leading order'' (LO), ``next-to-leading order'' (NLO),
and so forth. The NLO corrections for the spin-dependent cross sections
for inclusive-hadron and jet production were published
in~\cite{jssv,ddf} and~\cite{jsv,dfsv}, respectively. They are
crucial for making reliable quantitative predictions and for analyzing the 
forthcoming RHIC data in terms of spin-dependent parton densities.
The corrections can be sizable and they reduce the dependence on 
the factorization/renormalization scale $\mu$ in
Eq.~(\ref{eq:xsecfact}). In case of jet production, NLO corrections 
are also of particular importance since it is only at NLO that the 
QCD structure of the jet starts to play a role.

\begin{figure}[t]
\centerline{\epsfxsize=2.6in\epsfbox{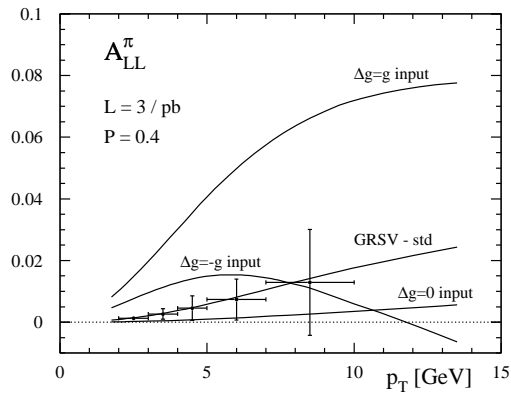}}   
\caption{NLO spin asymmetry$^{\;2}$ 
for $\pi^0$ production, using several GRSV
polarized parton densities$^{\;7}$ 
with different gluon polarizations. \label{fig1}}
\end{figure}

Figure~\ref{fig1} shows NLO predictions for the spin asymmetry 
$A_{\mathrm{LL}}$ for high-$p_T$ pion
production for collisions at $\sqrt{S}=200$~GeV at RHIC. 
We have used various sets of polarized parton densities of~\cite{grsv},
which mainly differ in $\Delta g$. As one can
see, the spin asymmetry strongly depends on $\Delta g$, 
so that measurements of $A_{\mathrm{LL}}$ at RHIC should give 
direct and clear information. The ``error bars'' in the figure
are uncertainties expected for measurements with an integrated luminosity 
of 3/pb and beam polarization P=0.4. We note that
PHENIX has already presented preliminary data~\cite{phenixall} for
$A_{\mathrm{LL}}$. We also mention that the figure shows that at lower $p_T$
the asymmetry is not sensitive to the {\it sign} of $\Delta g$. This is 
related to the dominance of the $gg$ scattering channel which is
approximately quadratic in $\Delta g$. In fact it can be 
shown that $A_{\mathrm{LL}}$ in leading-power QCD can hardly be
negative at $p_T$ of a few GeV~\cite{jksv}. One may obtain
better sensitivity to the sign of $\Delta g$ by expanding kinematics to the 
forward rapidity region.

\begin{figure}[ht]
\vspace*{-1cm}
\centerline{\epsfxsize=3.1in\epsfbox{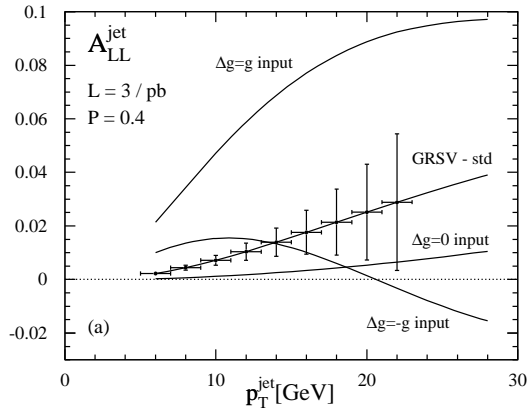}}   
\caption{Same as Fig.~1, but for inclusive jet production$^{\;3}$ 
at RHIC. \label{fig2}}
\end{figure}

Figure~\ref{fig2} shows  predictions for the spin asymmetry 
$A_{\mathrm{LL}}$ for high-$p_T$ jet production. The gross
features are rather similar to the pion asymmetry, except that 
everything is shifted by roughly a factor two in $p_T$. This is due 
to the fact that a pion takes only a certain fraction of $\sim{\mathcal{O}}
(50\%)$ of the outgoing parton's momentum, so that the hard scattering
took place at roughly twice the pion transverse momentum. 
A jet, however, will carry the full transverse momentum of a produced 
parton. 

We emphasize that PHENIX and STAR
have presented measurements~\cite{rhicpion} of the 
unpolarized cross section for $pp\to \pi^0 X$. These are well 
described by the corresponding NLO QCD 
calculations~\cite{jssv,ddf}, providing confidence 
that the NLO pQCD hard-scattering framework 
is indeed adequate in the RHIC domain. 
This is in contrast to what was found in 
comparisons~\cite{aur} between NLO theory
and data for inclusive-hadron production taken in the
fixed-target regime. We will turn to this issue next.

\section{Threshold resummation for inclusive-hadron production} 
One may further improve the theoretical 
calculations by an all-order resummation of large 
logarithmic corrections to the partonic cross sections~\cite{ddfv}.
At partonic threshold, when the initial partons have 
just enough energy to produce a high-transverse momentum 
parton (which subsequently fragments into the observed pion) 
and a massless recoiling jet, the phase space available for gluon 
bremsstrahlung vanishes, resulting in large logarithmic corrections to 
the partonic cross section. For the rapidity-integrated cross
section, partonic threshold is reached when $\hat{x}_T\equiv 
2 \hat{p}_T/\sqrt{\hat{s}}=1$,
where $\sqrt{\hat{s}}$ is the partonic center-of-mass (c.m.) energy, 
and $\hat{p}_T$ is the transverse momentum of the produced 
parton fragmenting into the hadron. 
The leading large contributions near threshold arise as $\alpha_s^k
\ln^{2k}\left(1-\hat{x}_T^2\right)$ at the $k$th order in 
perturbation theory. Sufficiently
close to threshold, the perturbative series will be only useful if
such terms are taken into account to all orders in $\alpha_s$, which is
achieved by threshold resummation~\cite{dyresum}. 
This resummation has been derived for a number of cases of
interest, to next-to-leading logarithmic (NLL) order, in particular
also for jet production~\cite{KS} which proceeds through 
the same partonic channels as inclusive-hadron production. 

The larger $\hat{x}_T$, the more dominant the threshold logarithms 
will be. Since $\hat{s}=x_a x_b S$, where $x_{a,b}$ are the partonic 
momentum fractions and $\sqrt{S}$ is the hadronic c.m. energy, 
and since the parton distribution functions
fall rapidly with increasing $x_{a,b}$, threshold effects 
become more and more relevant as the hadronic scaling variable 
$x_T\equiv 2 p_T/\sqrt{S}$ goes to one. This means that 
the fixed-target regime with 3~GeV $\lesssim p_T\lesssim$ 10~GeV
and $\sqrt{S}$ of 20$-$30~GeV is the place where threshold
resummations are expected to be particularly relevant and useful. 

\begin{figure}[ht]
\vspace*{-10mm}
\centerline{\epsfxsize=2.6in\epsfbox{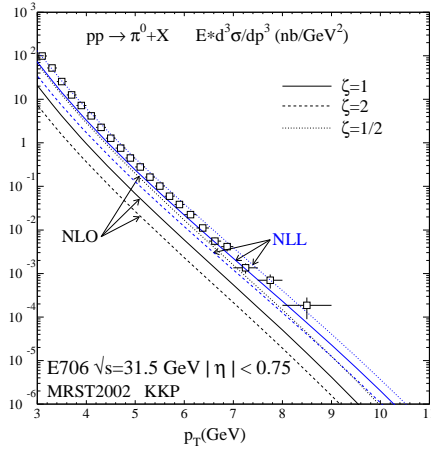}}   
\caption{NLO and NLL resummed$^{\;4}$ results for the cross
section for $pp\to \pi^0X$ for E706 kinematics. Results
are given for three different choices of scales, 
$\mu=\zeta p_T$, where $\zeta=1/2, 1, 2$. Data are
from$^{\;14}$. \label{fig3}}
\end{figure}

The resummation is performed in Mellin-$N$ moment space, where
the logarithms $\alpha_s^k\ln^{2k}\left(1-\hat{x}_T^2\right)$ 
turn into $\alpha_s^k \ln^{2k}(N)$, which then exponentiate.
For inclusive-hadron production, because of the color-structure of 
the underlying Born $2\to 2$ QCD processes, one actually
obtains a {\it sum} of exponentials in the resummed expression.
Details may be found in~\cite{ddfv}. Here, we only give a 
brief indication of the qualitative effects resulting from
resummation. For a given partonic channel $ab\to cd$,
the leading logarithms exponentiate in $N$ space as 
\begin{equation}
\label{eq:res1}
\hat{\sigma}^{(res)}_{ab\to cd} (N) \propto 
\exp\left[ \frac{\alpha_s}{\pi}\left(C_a+C_b+C_c-\frac{1}{2} C_d\right)
\ln^2 (N) \right] \; ,
\end{equation}
where
\begin{equation} 
C_g=C_A=N_c=3 \;, \;\;\;C_q=C_F=(N_c^2-1)/2N_c=4/3 \; .
\end{equation} 
This exponent is clearly positive for each of the 
partonic channels, which means that the soft-gluon effects will lead to an 
enhancement of the cross section. Indeed, as may be seen from
Fig.~\ref{fig3}, resummation dramatically increases the cross section 
in the fixed-target regime. The example we give is a comparison
of NLO and NLL resummed predictions at $\sqrt{S}=31.5$~GeV
with the data of E706~\cite{e706} at that energy. We have used the ``KKP''
set of pion fragmentation functions~\cite{kkp}, and the 
parton distributions of~\cite{mrst}. We finally note that the 
results shown in Fig.~\ref{fig3} are also interesting with respect to 
the size of power corrections to the cross section. Resummation may
actually suggest the structure of nonperturbative power corrections. For 
a recent study of this for single-inclusive cross sections, see~\cite{stwv}.

\section*{Acknowledgments}
I am grateful to D.\ de Florian, B.\ J\"{a}ger, S.\ Kretzer, 
A.\ Sch\"{a}fer, G.\ Sterman, and M.\ Stratmann for 
fruitful collaborations on the topics presented here. I thank RIKEN,
BNL and the U.S.\ DoE
(contract number DE-AC02-98CH10886) for
providing the facilities essential for the completion of this work.

\end{document}